\documentclass[11pt]{article}
\usepackage[a4paper, total={6in, 8in}]{geometry}
\usepackage{dblfloatfix}
\usepackage{xcolor}
\usepackage{lipsum}
\usepackage{amssymb}

\usepackage{mathtools,tikz}
\usetikzlibrary{arrows.meta}
\usepackage{caption}
\usepackage{subcaption}
\usepackage{float}
\usepackage{amsmath}
\usepackage{lastpage}
\usepackage{fancyhdr}
\usepackage{graphicx}
\usepackage{color,url}
\graphicspath{{C:/Users/conor/Pictures}}
\usepackage{commath,comment}
\begin{document}

\title{Long interfacial internal waves with currents: current profiles with quadratic depth-dependence }

\author{Conor T. Curtin and Rossen I. Ivanov \\[5pt]
{\sl School of Mathematics and Statistics, Technological University Dublin} \\
{\sl Grangegorman Lower, Dublin D07 ADY7, Ireland}\\[5pt]
}

\date{}
\maketitle

\begin{flushright}
 \begin{minipage}[h]{4.2in}
 \end{minipage}
\end{flushright}
\bigskip

\begin{abstract}
Internal waves often propagate in the presence of currents. Internal waves and currents are widespread in the world's oceans. The equatorial internal waves and the equatorial undercurrent (EUC) are perhaps the most representative illustration of this phenomenon. The current profile in this study is a quadratic function of depth, which is in agreement with the known results of the EUC. With this assumption the integrable nonlinear Gardner's equation is derived as a model for the wave motion of the thermocline, separating the two layers of slightly different densities. 
In addition, we derive the generalised Gardner's equation in the case of variable bottom. We demonstrate that the coefficients of the equation are very sensitive with respect to the parameters of the EUC.
\end{abstract}


\section{Introduction}

The dynamics of the upper-ocean involve many complex processes, including, for instance, the interaction between wind, waves, currents, and global circulation systems. Ocean currents are characterised as steady mean flows of water in a prevailing direction. The main types of ocean waves include surface waves (generated by wind), internal waves (due to density stratification), and lee waves (generated by bottom topography). Ocean subcurrents exist at depths beneath the surface that cause interplay with the wave motion in the fluid bulk. These phenomena drive global ocean circulation, nutrient mixing, and shape coastal environments. The dynamics of the wave-current interactions is essentially non-linear. 
\par
The surface wave dynamics in the irrotational case admits a beautiful Hamiltonian mathematical formulation, presented in the seminal work of Zakharov \cite{Z68}. Numerous extensions, in particular those in \cite{Craig1993,CraigGroves1} allowed for the tractable set-up using the Dirichlet-Neumann operator for the fluid domain. The effect of a linear shear current can be included in this formulation for the movement of two-dimensional fluids \cite{NearlyHamiltonian,W}. 
The Hamiltonian description for two superimposed irrotational fluid layers, modelling internal waves, has been presented in
a series of works, e.g. \cite{CGK,CGS}. Following these ideas, the Hamiltonian formulation has been extended for a system of surface and internal interfacial waves interacting with linearly sheared current in \cite{CIM-16,CI-19,Iv17}.  

Interfacial internal waves are formed at the thermocline - a thin fluid layer that separates the upper warmer layer of lower density and the lower cold layer of higher density. The density difference is mainly due to salinity. The two-dimentional motion of internal waves and currents is widely observed in the Equatorial Pacific area. The Pacific Ocean within a small latitudinal deviation from the equator possesses rich dynamics and structures \cite{Boyd,FB}. In particular, we see a strong and highly pronounced stratification, in which there is a small and relatively warm upper layer, followed by a deep layer of greater density (around 1\%) in difference. The EUC can be characterised as an ocean jet current, although its width is around 300km it is relatively fast flowing and appears to be highly depth dependent. It is uniquely situated in an equatorial trap, where due to net zero Coriolis effect, the EUC is forced to act in a narrow band unlike Eastern Boundary Currents. 
The EUC flows at depths of 50 m to approximately 300 m. The flow of the EUC is eastwards at between $1$ m/s and $1.5$ m/s, counter to the surface flow which is driven by the trade winds. The EUC is of particular importance due to its percieved interaction with the El-Ni\~no phenomenon. The EUC is comparatively colder than the surface sea temperature, thus its upwelling can bring cooler surface water temperatures. Although we know much about the generalities of the EUC, the effect the current has on the Earth's climate is not well understood. The EUC is of central importance to the mechanism of the El Ni\~no - La Ni\~na southern oscillation. The EUC travels the entirety of the Equatorial Pacific, where the bathymetry of the continental shelf forces the current to upwell around the Galapagos Islands. 

Some specific mathematical models for the motion of the thermocline in the Equatorial area under the influence of EUC 
assuming linear piece-wise shear are presented in \cite{CoIv2,CoIv3,CuIv,CuIv2,henry,IvIv} etc. However, the current profile of the EUC has a more complicated structure. Numerous attempts have been made to model the current, as discussed in short in \cite{Curt2}. In \cite{Mamaev} for example, a quartic polynomial formula was proposed for the EUC profile, which describes the flow reversal. In the Appendix 2 we briefly address this problem and justify the piece-wise quadratic current profile, shown in Fig. \ref{FigU}.  

In this study, we show that the quadratic profile of the EUC can be accommodated into the derivation of the nonlinear model equations for the thermocline. This leads to a complication related to the vorticity equation, since in the considered model the vorticity field is not constant. We resolve this problem, using the approach proposed in \cite{Curt2} for a single fluid layer. An alternative approach based on other approximations and the so-called Burns condition is given in \cite{GQ}.  A Higher order KdV model (HKdV) is derived in \cite{GPP} for a general form of the shear current. The derivation follows the continuous stratification formalism, which uses the stream function and does not deal directly with the vorticity equation. Since the HKdV models are not always integrable the connection to the integrable equations of this type is given by some specific transforms (Kodama transforms), see for example \cite{henry}.  

Recently in \cite{Va} a method which gives complete methodology for internal wave detection is proposed. The method is divided into three main stages: input data pre-processing, parameter extraction and modeling. For the modelling part a KdV (Korteweg–de Vries) solver is used. KdV solver takes the internal wave parameters as input and it gives the velocity, density plots of internal waves. Therefore, for this type of analysis, the use of the correct type of model with all necessary parameters is essential.  

In Section \ref{sec:2} we outline the setup of the problem, introduce the governing equations, and the main assumption for the velocity field decomposition involving the current profile. 

In Section \ref{sec:3} we introduce the mathematical objects called Dirichlet-Neumann Operators (DNO), which play an important role in our further analysis. We express the kinematic boundary condition on the interface in terms of the DNO.

In Section \ref{sec:4} we express the Bernoulli-type equation for the motion of the interface in terms of the DNO and the vortivity-related conditions for the interface. 

In Section \ref{sec:5} we derive an asymptotic model equation equivalent to the KdV equation \cite{KdV} in the case on the Bousinesq scaling. Furthermore in the special case of vanishing quadratic nonlinear term in the KdV model we derive the higher order Gardner's equation which includes a cubic nonlinearity term and is also an integrable nonlinear model in Section \ref{sec:Gardner}. 

In Section \ref{sec:7} we show the impact of the current and its parameters on the solitary waves. It turns out that the solitons are very sensitive with respect to these parameters.

  In Section \ref{sec:8} we extend Gardner's equation in the case of variable bottom. 
  
\section{Setup, preliminaries and assumptions}\label{sec:2}
\begin{figure}
\begin{center}
    
\begin{tikzpicture}[x=0.8cm, y=0.8cm]

\def\L{12}           
\def\hOne{2.5}       
\def\hBot{4}       
\def\A{1.0}          
\pgfmathsetmacro\k{2*pi/1.5*\L} 
\pgfmathsetmacro\xShear{5.3}    
\pgfmathsetmacro\ShearScale{2.0}

\pgfmathdeclarefunction{eta}{1}{\pgfmathparse{\A*sin(\k*#1)}}

\pgfmathdeclarefunction{Uxcurve}{1}{%
  \pgfmathparse{\xShear - \ShearScale*((#1/\hOne)^2)}%
}

\draw[very thick] (-\L/2,  \hOne) -- (\L/2,  \hOne);  
\draw[very thick] (-\L/2, -\hBot) -- (\L/2, -\hBot);  

\node[left] at (-\L/2 - 0.45,  \hOne) {$z = h_1$};
\node[left] at (-\L/2 - 0.45, -\hBot) {$z = -h$};

\draw[->] (-\L/2, -\hBot - 0.7) -- (\L/2, -\hBot - 0.7) node[right] {$x$};


\begin{scope}
  \clip (-\L/2,-\hBot) rectangle (\L/2,\hOne);
  \fill[blue!3]
    (-\L/2, \hOne) -- (\L/2, \hOne) --
    plot[domain=\L/2:-\L/2, samples=200] (\x, {eta(\x)}) -- cycle;
  \fill[blue!10]
    (-\L/2,-\hBot) --
    plot[domain=-\L/2:\L/2, samples=200] (\x, {eta(\x)}) --
    (\L/2,-\hBot) -- cycle;
\end{scope}

\draw[very thick, blue!70!black]
  plot[domain=-\L/2:\L/2, samples=400] (\x, {eta(\x)});

\draw[loosely dashed] (-\L/2, 0,0) -- (\L/2, 0,0);
\node[left] at (-\L/2 -0.45, 0) {$z=0$};

\node at (0,  1.35) {\text{The domain }\, $\Omega_1 (\eta), \,\,\, \text{density} \, \rho_1$};
\node at (0, -1.35) {\text{The domain }\, $\Omega (\eta), \,\,\, \text{density} \, \rho$};

\end{tikzpicture}
\caption{Schematic of a two dimensional internal wave model. The two domains are separated by the interface $z=\eta(x,t).$ The densities are $\rho > \rho_1.$}
\label{fig:2layer}
\end{center}
\end{figure}
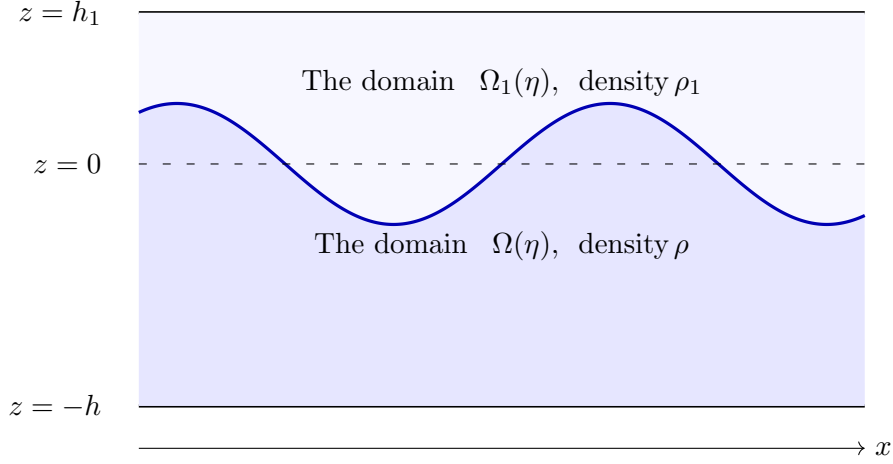

We introduce a two layer fluid model in the $(x,z)$ Cartesian plane, which is bounded above by a flat surface (``rigid-lid'') at depth $z=h_1$ and below by a flat bottom at $z=-h.$  We assume a density stratification such that there exist wave motion at the thermocline, whose elevation is given by $z=\eta(x,t)$. The equilibrium of the thermocline is at $z=\eta(x,t)=0,$ therefore,  
\begin{equation}
\int_{\mathbb{R}} \eta(x',t) \, dx'=0.
\end{equation}  
Let the domain $\Omega:=\{(x,z), \, -h <z < \eta(x,t)  \}$ be the bottom layer with constant density $\rho$ and 
$\Omega_1:=\{(x,z), \,  \eta(x,t)<z<h_1  \}$ be the upper layer with constant density $\rho_1$; and assume that $\rho > \rho_1$, see Fig. \ref{fig:2layer}.

\subsection{Governing equations}

The fluid is governed by the incompressibility condition of the velocity field $\vec{V}(x,z,t):=(u,w)$ and $\vec{V_1}(x,z,t):=(u_1,w_1)$ in each layer
\begin{equation}
\text{div} \vec{V}=0 = \text{div} \vec{V_1},
\end{equation}
and the dynamics of the fluid are governed by Euler's equations for each layer with subscript $"1"$ denoting the upper layer:
\begin{equation} \label{EU0}
    \begin{split}
&u_t+uu_x+wu_z=-\frac{1}{\rho}P_x \\
&w_t+uw_x+ww_z=-\frac{1}{\rho}P_z -g
\end{split}
\end{equation}
in the domain $\Omega$ and
\begin{align} \label{EU1}
\begin{split}
&u_{1,t}+u_1u_{1,x}+w_1u_{1,z}=-\frac{1}{\rho_1}P_{1,x} \\
&w_{1,t}+u_1w_{1,x}+w_1w_{1,z}=-\frac{1}{\rho_1}P_{1,z}-g
\end{split}
\end{align}
in $\Omega_1$. The Earth's acceleration is denoted by $g,$ and $P, P_1$ is the pressure in the corresponding domain. The Coriolis parameter at the equator is zero, so the Coriolis forces have a negligible effect on the equatorial fluid motion - see Appendix 1 for the details. Otherwise, these forces play an important role in the formation of a two-dimensional fluid flow in the vertical plane along the equator, see for example \cite{RJ17}.

\begin{figure}
\begin{center}
\begin{tikzpicture}[>=Stealth, thick, scale=0.95]

\def\Umin{-0.2}
\def\Umax{1.2}

\def\xleft{-0.3}   
\def\xright{3.2}   
\def\xext{-1.3}    

\pgfmathsetmacro{\Uxscale}{(\xright-\xleft)/(\Umax-\Umin)}

\newcommand{\UtoX}[1]{\xleft + \Uxscale*((#1)-\Umin)}

\draw[->] (\xleft, -4.5) -- (\xleft, 3.0) node[above, font=\small] {$z$};
\draw[->] (\xext,  -4.5) -- (4.5,  -4.5) node[right, font=\small] {$U(z)$};

\draw[thick] (\xext,  1.2) -- (4.2,  1.2);   


\fill[blue!20]
  (\xext, 1.2) --     
  (\xleft, 1.2) --    
  (\xleft, 0.6) --    
  cycle;

\node[left, font=\footnotesize, blue!70!black]
  at (\xext, 0.9) {flow reversal};

\draw[dashed, blue!70]
  (\xext, 0.6) -- (4.2, 0.6);

\fill[orange!25]
  (\xleft, -2.8) --
  plot[domain=-2.8:0.8, samples=60, variable=\z]
    ({\UtoX{-0.2 + 1.4*(1 - ((\z+1.0)/1.8)^2)}}, \z) --
  (\xleft, 0.8) -- cycle;

\draw[orange!80!red, very thick]
  plot[domain=-2.8:0.8, samples=60, variable=\z]
    ({\UtoX{-0.2 + 1.4*(1 - ((\z+1.0)/1.8)^2)}}, \z);

\draw[orange!80!red, very thick]
  (\xleft, -2.8) -- (\xleft, 0.8);

\fill[red!70!black] ({\UtoX{1.2}}, -1.0) circle (3pt);
\draw[dashed, red!50] (\xleft, -1.0) -- ({\UtoX{1.2}}, -1.0);

\draw (-0.45, -1.0) -- (-0.15, -1.0);
\node[left, font=\footnotesize, red!70!black] at (-0.45, -1.0) {$z=0$};

\draw (-0.45, -2.8) -- (-0.15, -2.8);
\node[left, font=\footnotesize] at (-0.45, -2.8) {$z=-d_0$};

\draw (-0.45, 1.2) -- (-0.15, 1.2);
\node[left, font=\footnotesize] at (-0.45, 1.4) {$z=h_1$};

\draw ({\UtoX{-0.2}}, -4.65) -- ({\UtoX{-0.2}}, -4.35);

\draw ({\UtoX{1.2}}, -4.65) -- ({\UtoX{1.2}}, -4.35);
\node[below, font=\footnotesize, red!70!black]
  at ({\UtoX{1.2}}, -4.65) {$U_{max}$};

 \draw ({\UtoX{-0.5}}, -4.65) -- ({\UtoX{-0.5}}, -4.35);
\node[below, font=\footnotesize, red!70!black]
  at ({\UtoX{-0.5}}, -4.65) {$U^*$};

\draw[blue!80!black, ->, very thick]
  ({\UtoX{-0.05}}, 0.9) -- ({\UtoX{-0.25}}, 0.9);

\draw[orange!80!red, ->, very thick]
  ({\UtoX{0.05}}, -1.5) -- ({\UtoX{0.5}}, -1.5);
\node[font=\small\bfseries] at (2.0, 2.5) {};

\end{tikzpicture}
\caption{The quadratic flow profile of the EUC in the region of interest. The flow reverses close to the surface in reality due to wind shear in the westerly direction.}
\label{FigU}
\end{center}
\end{figure}
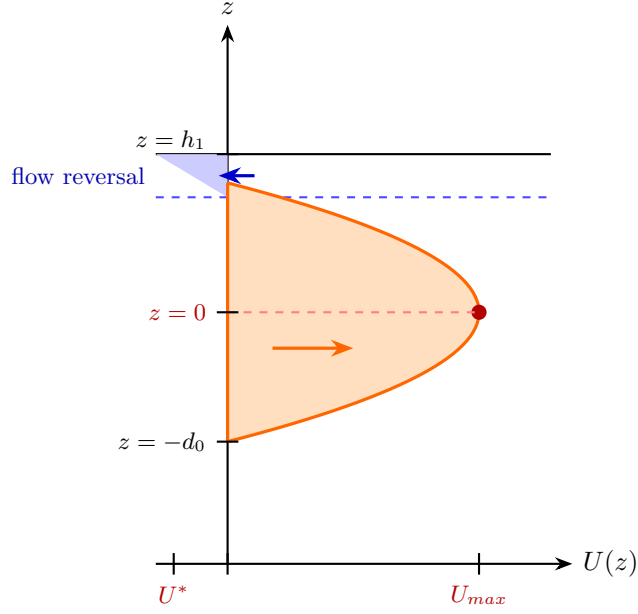

\subsection{Velocity field decomposition}
The case with linear shear has been extensively studied, \cite{NearlyHamiltonian,CIM-16}. Following these works, we introduce
velocity vector fields whose horizontal component $u$ depends on a shear current of the form $U(z)=\kappa+\gamma z + \beta z^2,$ that is,
\begin{align}\label{u}
       u(x,z,t)& = \varphi_x (x,z,t) + U(z)=  \varphi_x (x,z,t)+\kappa+\gamma z + \beta z^2 = \psi_z(x,z,t), \\
       w(x,z,t) &=\varphi_z (x,z,t)=-\psi_x(x,z,t), \label{w}
\end{align}
where $\kappa$, $\gamma$ and $\beta$ are constant parameters characterizing the current and $\psi$ is the corresponding stream function. Similarly, for the upper layer, we introduce the analogous quantities supplemented by a subindex ``1''.
\begin{align}\label{u1}
    u_1(x,z,t)&= \varphi_{1,x} (x,z,t) + U_1(z)=  \varphi_x (x,z,t)+\kappa_1+\gamma_1 z + \beta_1 z^2=\psi_{1,z}(x,z,t), \\
    w_1(x,z,t)&=\varphi_{1,z} (x,z,t)=-\psi_{1,x}(x,z,t). \label{w1}
\end{align}
The linear part $\kappa+\gamma z $  introduces constant vorticity, which automatically satisfies the vorticity equation. So we will examine separately the input of the quadratic term in the current. Let us for simplicity analyse the lower layer first.  The boundary of the layer is the interface (thermocline) $z=\eta(x,t).$

We observe that the incompressibility condition is automatically satisfied if $\varphi$ is harmonic function, since 
\begin{equation}
    \text{div} \vec{V}=u_x+w_z= \varphi_{xx}+ \varphi_{zz} .
\end{equation}  
Actually, the relation between $\varphi$ and $\psi$ is given through the Cauchy-Riemann equations 
\begin{equation} \label{CR}
    \varphi_x=\Psi_z, \quad \varphi_z=-\Psi_x
\end{equation}
for the related quantity
\begin{equation}
    \Psi:= \psi - \left(\kappa z+ \frac{1}{2}\gamma z^2 + \frac{1}{3} \beta {z^3}\right).
\end{equation} Clearly, similar Cauchy-Riemann equations hold for the quantities of the upper layer as well. This allows for an analytic continuation of the complex potentials $\varphi+i \Psi, $ $\varphi_1+i \Psi_1, $ of the complex variable $x+iz$ from their boundary values to the  interior of the fluid domains $\Omega$ and $\Omega_1.$

\begin{figure}[ht!]
    \centering
\includegraphics[scale=0.50]{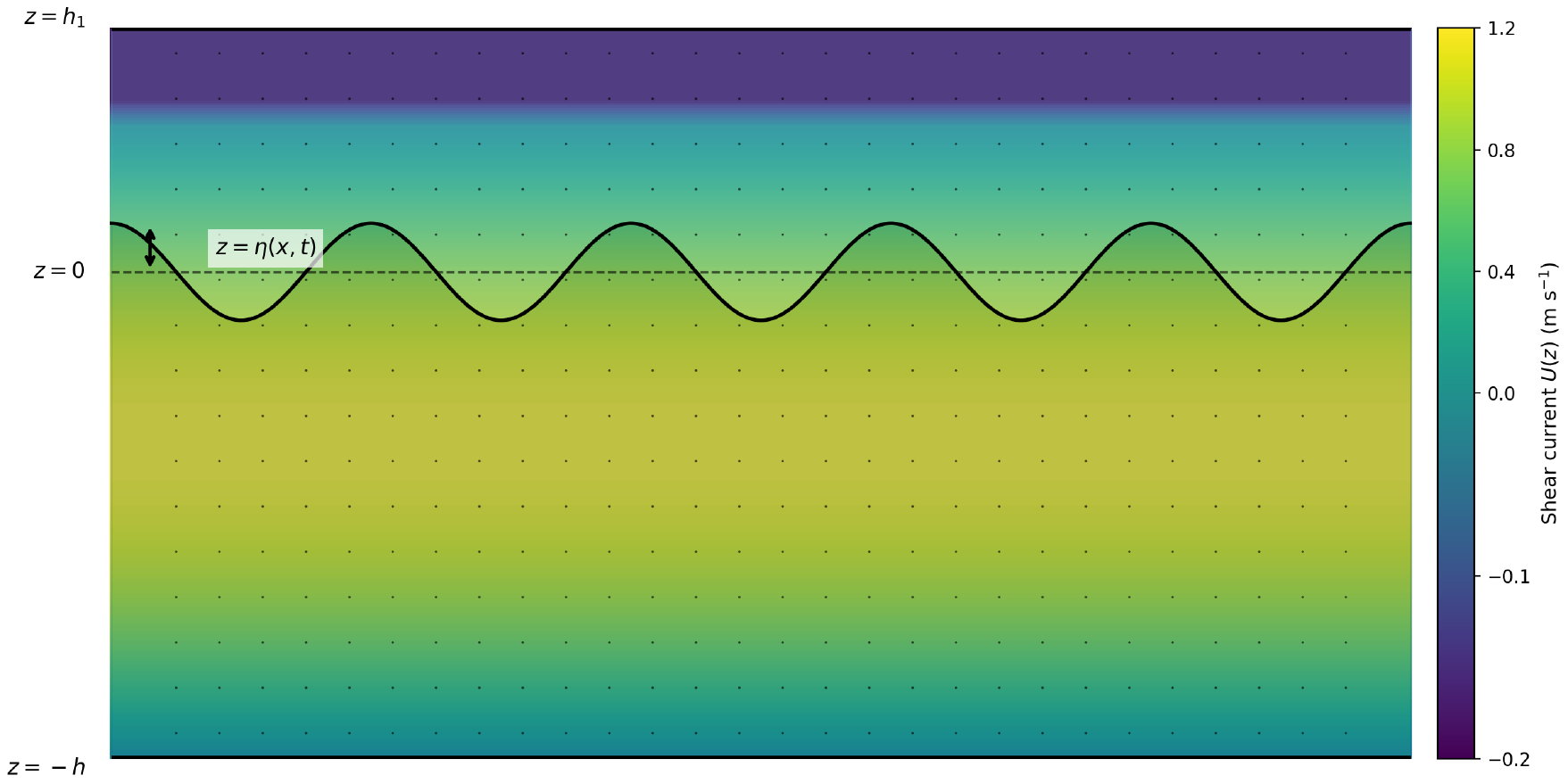}
\caption{Internal waves in the presence of shear current. }
\label{Fig-WCI}
\end{figure}

The quadratic current profile is physically justified in the Appendix 2, see also the sketch in Fig. \ref{FigU}.  
The wave motion in the presence of current is illustrated schematically in Fig. \ref{Fig-WCI}.

\section{Dirichlet - Neumann Operators and Kinematic Boundary Conditions} \label{sec:3}

The two potentials in \eqref{u} -- \eqref{w}  and \eqref{u1} -- \eqref{w1}  are harmonic functions in their respective domains, $\Omega$ and $\Omega_1,$
\begin{equation}
\nabla^2 \varphi=0=\nabla^2 \varphi_1.
\end{equation}
The domain of the fluid is characterised by the boundary conditions:
\begin{align}
\begin{cases}
\varphi_z(x,-h)=0 \\
\varphi_{1,z}(x,h_1)=0
\end{cases}
\end{align}
where the flat bed is at $z=-h$ and the flat surface (rigid lid) is at $z=h_1$. These boundary conditions guarantee that no fluid particle will pass through the boundary, i.e. the normal velocity to the bottom and top of the domain is zero respectively. The unperturbed interface of $\Omega$ and $\Omega_1$ is at $z=0$. \\
These boundary conditions are accompanied by the kinematic boundary condition at the interface, where subscript ``$c$'' denotes evaluation at $z=\eta(x,t)$:
\begin{align}\label{kin}
\begin{cases}
\eta_t= -\eta_x((\varphi_x)_c)+U(\eta)+(\varphi_z)_c \\
\eta_t= -\eta_x((\varphi_{1,x})_c)+U_1(\eta)+(\varphi_{1,z})_c.
\end{cases}
\end{align}
$U(\eta)$ and $U_1(\eta)$ are the respective shear currents in each layer evaluated on the wave. 

We point out that the tangential component of the velocity field with respect to the thermocline is not continuous. This also is the case when the fluid motion is irrotational. However, for the purposes of understanding the propagation of interfacial waves, we can assume that any instabilities at the interface (due to the tangent velocity discontinuity) do not significantly affect the wave dynamics provided the waves are long compared to the distance over which the interface thickens due to any potential mixing, see the discussion in \cite{Suth}.

The Bernoulli equation and the interface and the kinematic boundary condition together will describe in full the dynamics of the wave-current interactions on the interface. Thus in the following sections we seek to derive the Bernoulli equation from  \eqref{EU0},\eqref{EU1} and to write it together with \eqref{kin} in terms of canonical variables in the Hamiltonian framework. We first introduce values of the potentials on the interface $z=\eta(x,t)$ :
\begin{equation}\label{short}
\begin{cases}
 \phi(x,t):= \varphi(x,\eta(x,t),t) = (\varphi)_c \\
\phi_1(x,t):= \varphi_1(x,\eta(x,t),t)  = (\varphi_1)_c . 
\end{cases}
\end{equation}
The expressions in \eqref{short} will be used interchangeably so as to aid the formulation and allow us to define the canonical momentum given by 
\begin{equation}\label{xi}
\xi(x,t) = \rho (\varphi)_c - \rho_1 (\varphi_1)_c = \rho\phi - \rho_1\phi_1.
\end{equation}
We assume also that $\eta(\cdot,t),$ $\xi(\cdot,t) $ and the potentials $ \phi(\cdot,t),$ $ \phi_1(\cdot,t)$ are in the Schwartz space of rapidly decreasing functions  $\mathcal{S}(\mathbb{R})$ for any $t$. This, of course, implies that these functions decay to zero as $x \rightarrow \pm  \infty.$

The so called Dirichlet-Neumann operators (DNO) are very important for the formulation of the dynamic equations. 
These are defined in each layer by 
\begin{equation}
G(\eta) \phi = \left(\frac{ \partial \varphi}{\partial \textbf{n}} \right)_c \cdot \sqrt{1+\eta_x^2}
\end{equation}
and 
\begin{equation}
G_1 (\eta) \phi_1 = \left(\frac{ \partial \varphi_1}{\partial \textbf{n}_1} \right)_c \cdot \sqrt{1+\eta_x^2}
\end{equation}
where 
\begin{align*}
 \textbf{n} = \frac{(- \eta_x,1)}{\sqrt{1+\eta_x^2}} \, \,\, \text{and}\, \, \,\textbf{n}_1= \frac{(\eta_x,-1)}{ \sqrt{1+\eta_x^2}}=-\textbf{n}
\end{align*}
are the respective outer unit normal's in each layer. Hence we may admit the following expressions 
\begin{equation}
\begin{cases}
G(\eta) \phi  = -\eta_x (\varphi_x)_c + (\varphi_z)_c \\
G_1(\eta) \phi_1 = \eta _x(\varphi_{1,x})_c -( \varphi_{1,z})_c.
\end{cases}
\end{equation}
and from \eqref{kin} we may conclude that 
\begin{equation}\label{dno eta}
\eta_t = G(\eta)\phi - (\kappa +\gamma \eta + \beta \eta^2)\eta_x = -G_1(\eta) \phi_1 -(\kappa_1 + \gamma_1 \eta +\beta_1 \eta^2)\eta_x
\end{equation}
and therefore 
\begin{equation}
    G(\eta)\phi +G_1(\eta)\phi_1 = \mu
\end{equation}
where \begin{equation}\label{mu2}
\mu:= \left[(\kappa - \kappa_1) + (\gamma-\gamma_1)\eta +(\beta-\beta_1)\eta^2\right]\eta_x,
\end{equation} is explicitly given in terms of $\eta.$
We also define the operator 
\begin{equation}\label{DNO2}
B= \rho_1 G(\eta) + \rho G_1(\eta)
\end{equation} and with \eqref{DNO2} and \eqref{xi} we have 
\begin{equation}
    B \phi = \rho_1 G(\eta)\phi + \rho G_1(\eta)\phi =\rho_1 (\mu -G_1 \phi_1)+\rho G_1(\eta)\phi= \rho_1 \mu+ G_1 \xi.
\end{equation} from where we obtain $\phi.$ Similarly, we can express $\phi_1,$ thus we obtain expressions for $\phi$ and $\phi_1$ in terms of $\eta$ and $\xi,$ with the help of the the DNO's: 
\begin{equation}\label{dnob}
\begin{cases}
\phi(x,t) = B^{-1}\left(\rho_1 \mu +G_1\xi \right)=:\Phi[\xi(x,t), \eta(x,t)] \\
\phi_1(x,t) = B^{-1}\left(\rho \mu - G \xi \right)=: \Phi_1[\xi(x,t), \eta(x,t)].
\end{cases}
\end{equation}
The invertibility of the DNO and by extension it's linear combinations is discussed rigorously in \cite{Lan}. It is important to note that operators $G,G_1$ and $B$ are self-adjoint along with the operator $GB^{-1}G_1$. Therefore it is also true that $GB^{-1}G_1 =G_1B^{-1}G$. 

Let us assume $\kappa=\kappa_1,$ which insures the continuity of the current profile at rest when $\eta \equiv 0.$
Then, the kinematic boundary condition in terms of $\xi$ and $\eta$ can be written as
\begin{equation} \label{KBC1}
    \eta_t +U(\eta) \eta_x = \rho_1 G B^{-1} [((\gamma-\gamma_1) \eta + (\beta - \beta_1) \eta^2)\eta_x] + G B^{-1} G_1 \xi =: F[\xi, \eta],
\end{equation}
and equivalently,
\begin{equation} \label{KBC2}
     \eta_t +U_1 (\eta) \eta_x = -\rho G_1 B^{-1} [((\gamma-\gamma_1) \eta + (\beta - \beta_1) \eta^2)\eta_x] + G B^{-1} G_1 \xi =: F_1[\xi, \eta].
\end{equation}

The equivalent expressions \eqref{KBC1} and \eqref{KBC2} generalise the corresponding equations for linear shear currents from \cite{henry}. Note that these equations are written in maximally general form and are applicable in all propagation regimes. In what follows, many of the derivations will be valid only in the long wave regime and for this reason we need to introduce the physical scales.

\section{Bernoulli's equation and the vorticity equation}\label{sec:4}

Now  our aim is recasting Euler's equations in terms of potentials $\varphi$, $\varphi_1$ and stream-functions $\psi$, $\psi_1$ for each layer, which will return Bernoulli's equation for the motion when evaluated at the interface. 


In the lower domain $\Omega$ from \eqref{EU0} we obtain the equations
\begin{equation}\label{b1}
\begin{cases}
\left[ \varphi_t + \frac{\varphi_x^2+\varphi_z^2}{2} + U \varphi_x - U'(z) \psi +\frac{P}{\rho}+gz \right]_x = 0,  \\
\left[ \varphi_t + \frac{\varphi_x^2+\varphi_z^2}{2} + U(z) \varphi_x  - U'(z) \psi +\frac{P}{\rho}+gz  \right]_z +U(z)U'(z) + \psi U''(z)=0.
\end{cases}
\end{equation}
and we have 
\begin{align}\label{b2}
\varphi_t + \frac{\varphi_x^2+\varphi_z^2}{2} & + U \varphi_x - U'(z) \psi +\frac{P}{\rho}+gz +\frac{1}{2}U^2(z)=\tilde{F}(z),\\
 \tilde{F}'(z)& + U''(z) \psi=0, \label{ve}
\end{align} where $\tilde{F}(z)$ is a yet unknown function. Equation \eqref{ve} is the compatibility condition of the two equations in \eqref{b1}, which is equivalent to the vorticity equation. It is clear, that \eqref{ve} cannot be satisfied in the entire domain $\Omega.$ However, for the Bernoulli equation we need to satisfy all equations only on the interface, that is,
\begin{align}\label{b3}
(\varphi_t)_c + \frac{(\varphi_x^2)_c+(\varphi_z^2)_c}{2} & + U(\eta) (\varphi_x)_c - U'(\eta) (\psi)_c +\frac{P_c}{\rho}+g\eta +\frac{1}{2}U^2(\eta)=\tilde{F}(\eta),\\
 \tilde{F}'(\eta)& + U''(\eta) (\psi)_c=0, \label{ve2}
\end{align} and similarly, for the domain $\Omega_1$
\begin{align}\nonumber
(\varphi_{1,t})_c + \frac{(\varphi_{1,x}^2)_c+(\varphi_{1,z}^2)_c}{2} & + U_1(\eta) (\varphi_{1,x})_c - U'_1(\eta) (\psi_1)_c +\frac{P_{1,c}}{\rho_1}+g\eta +\frac{1}{2}U_1^2(\eta)=\tilde{F}_1(\eta),\\
 \tilde{F}_1'(\eta)& + U_1''(\eta) (\psi_1)_{c}=0, \nonumber
 \end{align}
On the interface the pressures are equal, $P_{1,c}=P_c$ and also $(\psi)_c=(\psi_1)_c $ thus we obtain the Bernoulli equation, supplemented with the two equations for the functions $\tilde{F}$ and $\tilde{F}_1:$
\begin{align}
  (\rho \varphi_t - \rho_1\varphi_{1,t})_c & + \frac{\rho (\nabla \varphi)^2_c-\rho_1 (\nabla \varphi_{1,c})^2_c}{2}-(\rho U'(\eta)-\rho_1 U' _1(\eta) )(\psi)_c +(\rho-\rho_1)g \eta \nonumber \\
  +&\rho U(\eta) (\varphi_x)_c - \rho_1 U_1(\eta)(\varphi_{1,x})_c + \frac{\rho U^2(\eta)- \rho_1 U_1^2(\eta)}{2}=\rho \tilde{F}(\eta) - \rho_1 \tilde{F}_1 (\eta),  
  \end{align}
In addition, we point out that the kinematic boundary conditions \eqref{kin} written in terms of the stream functions lead to the expression \begin{equation}
    \frac{d}{dx} \psi(x,\eta,t)= (\psi_x)_c+ (\psi_z)_c \eta_x = -\eta_t,
\end{equation}
 therefore  
\begin{equation}
    (\psi)_c = (\psi_1)_c=-\partial_x^{-1} {\eta}_t.
\end{equation} 

The potential derivatives can be expressed as functionals of $\xi$ and $\eta,$ see for example \cite{henry}
\begin{align}\label{allphi}
    ({\varphi}_x)_c&=\frac{\Phi_x - \eta_x F }{1+\eta_x^2}, \quad ({\varphi}_z)_c=\frac{F+ \eta_x\Phi_x }{1+\eta_x^2}, \\
     ({\varphi}_{1,x})_c&=\frac{\Phi_{1,x} - \eta_x F_1 }{1+\eta_x^2}, \quad ({\varphi}_{1,z})_c=\frac{F_1+ \eta_x\Phi_{1,x} }{1+\eta_x^2}. 
\end{align}
and therefore, 
\begin{align} 
    \rho (\varphi_t)_c -  \rho_1 (\varphi_{1,t})_c &= ( \rho \phi-  \rho_1 \phi_1)_t -\eta_t [ \rho (\varphi_z)_c -  \rho_1 (\varphi_{1,z})_c  ]   =\xi_t -\eta_t \frac{\rho F - \rho_1 F_1 + \eta_x \xi_x}{1+\eta_x^2}, \nonumber
\end{align}
and finally, the Bernoulli equation is

\begin{align} 
   & \xi_t -\eta_t \frac{\rho F - \rho_1 F_1 + \eta_x \xi_x}{1+\eta_x^2}+ \frac{\rho \Phi_x^2-\rho_1 \Phi_{1,x}^2+\rho F^2 - \rho_1 F_1^2}{2(1+\eta_x^2)}+ (\rho-\rho_1)g \eta \nonumber \\
    &+\frac{\rho U(\eta)(\Phi_x-\eta_x F)}{1+ \eta_x^2}-\frac{\rho_1 U_1( \eta)(\Phi_{1,x}-\eta_x F_1)}{1+ \eta_x^2}+ 
(\rho U'(\eta)-\rho_1 U' _1(\eta) )  \partial_x ^{-1}\eta_t  \nonumber \\
      & + \frac{\rho U^2(\eta)- \rho_1 U_1^2(\eta)}{2}=\rho \tilde{F}(\eta) - \rho_1 \tilde{F}_1 (\eta),  \label{Bernoulli}\\
  & \tilde{F}'(\eta) + 2\beta (\psi)_c=0, \quad \tilde{F}_1'(\eta) +2\beta_1  (\psi_1)_{c}=0. \label{ve12}
  \end{align}

Equations \eqref{ve12} are a reincarnation of the vorticity equation on the two sides of the interface. This may look odd, however, it is due to the fact that there is always a discontinuity of the vorticity at the interface (vortex sheet). This happens even in the case of irrotational flow in the two layers - see the comment in \cite{Suth}: ``  
The existence of vorticity does not contradict our original assumption that each
layer is irrotational. The vorticity predicted by the model is confined to the interface
and does not extend into the interior of each layer. Vorticity is generated at the
interface through the action of baroclinic torques where the density and pressure
gradients are misaligned.''

\section{Physical Scales \& Approximation} \label{sec:5}

The asymptotic expansions of the quantities of the governing equations necessitate a suitable nondimensionalisation and scaling of the variables with respect to some scale parameters. We consider a situation where the two depths $h$ and $h_1$ are of the same order of magnitude, so we can introduce a depth scale $ \mathfrak {h}$ such that $h,h_1= \mathcal O(\mathfrak{h})$. And, assuming that the amplitude of the internal waves is of typical order $ \mathfrak {a},$ we introduce the scaling parameters 
\begin{equation}\label{param}
\varepsilon= \mathfrak {a}/ \mathfrak {h}, \quad  \delta= k\mathfrak {h}, 
\end{equation}
for wavenumber $k=2\pi/\lambda$, where $\varepsilon$  measures the amplitude size relative to the depth of the layers, while $\delta$ measures the depth relative to the wavelength $\lambda$. We assume both $\varepsilon \ll 1$ and $\delta  \ll 1$, $\varepsilon \ll 1 $ being the assumption for small wave amplitudes, and $\delta \ll 1$ the consideration of large wave length (small depth).  These parameters can be used to nondimensionalise all quantities (cf. \cite{Johnson,CJ2})
\begin{align}
\bar{ h} &=  \mathfrak{h} h ,  \quad  \bar{ h}_1 =  \mathfrak{h} h_1 ,  \quad \bar{x}= \frac{\mathfrak{h}}{\delta} x, \quad \bar{y}= \frac{\mathfrak{h}}{\delta^2} y ,\quad \bar{t} = \frac{\mathfrak{h} }{\delta \sqrt{\bar{g}\mathfrak{h}}}  t, \quad \bar{\eta} = \epsilon \mathfrak{h} \eta,  \nonumber \\
\bar{\varphi}&= \frac{\epsilon}{\delta } \mathfrak{h } \sqrt{\bar{g}\mathfrak{h}} \varphi, \quad 
 \quad  \bar{\varphi}_1= \frac{\epsilon}{\delta } \mathfrak{h } \sqrt{\bar{g} \mathfrak{h}} \varphi_1. 
\end{align}
The dimensional quantities are marked with a bar, while the non-dimensional ones are without a bar. 
The governing equations may be written of course in non-dimensional form, hence we seek equations which depend only on the non-dimensional variables (assumed to be of order 1) and the scale factor $\varepsilon$ and $\delta$.
We assume that the non-dimensional values of $\kappa,$ $ \gamma, $ $\gamma_1, $ $\beta, $ $\beta_1$ are of order 1. The non-dimensional version of Earth's acceleration $\bar{g}=9.81$ m/s$^2$ is  $g=1$, but we will retain $g$ throughout the following in order to easily revert back to the dimensional form of the equations.  

In the leading order, the linear equations have $\sin$ and $\cos$ solutions; therefore, the $x$-derivative acts on a monochromatic wave $\exp{i(kx-\omega(k) t})$ through $ik$ multiplication. The wavenumber is defined as $k=\frac{2\pi}{\lambda}$ thus $kh=2\pi \delta \sim \delta$ implying that the action of $\partial_x$ on the quantities brings a factor of $ \delta $.  The wave equation (which is the model in the leading order) necessitates both the $x-$ and $t-$derivatives to be of the same order, thus $\partial_t$ also brings a factor of $ \delta. $  

Under different scaling regimes, the order of $\varepsilon$ and $\delta$ will be related. First, we will study the Boussinesq regime, for which $\mathcal{O}(\varepsilon)= \mathcal{O}(\delta^2)$. 
As a special case we then study the case of higher order KdV (HKDV)  for which $\mathcal{O}(\varepsilon)= \mathcal{O}(\delta)$.  The leading idea is to use asymptotic expansions for the physical quantities in terms of $\varepsilon$ and $\delta,$ see for example also \cite{Bona,henry}.


Both $G(\eta) $ and $G_1(\eta)$ may be expanded as asymptotic series in $\varepsilon$ and $\delta$, for details we refer to \cite{Lan}, \cite{CraigGroves1}, \cite{Craig1993}. 
\begin{equation}\label{dnog}
    G(\eta)= -(\delta^2) h\partial_{x}^2-(\delta^4)\frac{1}{3} h^3 \partial_{{x}} ^4 -(\delta^2\varepsilon) \partial_{{x}} {\eta} \partial_{{x}} - (\delta^4\varepsilon )\partial_{{x}}^2 {\eta} \partial_{{x}}^2 + \delta^6 \frac{2h^2}{15}\partial^6_{{x}}+ ... ,
   \end{equation}
and with equivalent expression for $G_1(\eta)$ 
\begin{equation}\label{dnog1}
     G_1(\eta)= -(\delta^2) h_1\partial_{x}^2-(\delta^4)\frac{1}{3} h_1^3 \partial_{{x}} ^4 +(\delta^2\varepsilon) \partial_{{x}} {\eta} \partial_{{x}} + (\delta^4\varepsilon )\partial_{{x}}^2 {\eta} \partial_{{x}}^2 + \delta^6 \frac{2h_1^2}{15}\partial^6_{{x}}+ ... .
    \end{equation}

\subsection{KdV Scaling Regime}\label{sec:kdv}
For simplicity, we consider first the so-called KdV (or Boussinesq) regime where  $\mathcal{O}(\varepsilon)=\mathcal{O}(\delta^2).$  The quadratic part of the current does not affect the linear propagation regime in leading order. Then $\eta$ satisfies an equation of the form \cite{CoIv2,henry}
\begin{align} \label{c0}
        \eta_t& + c_0 \eta_x  = \mathcal{O}(\varepsilon), \quad c_0 = \kappa +  \frac{1}{2}\left( - \alpha_1 \Gamma \pm \sqrt{ \alpha_1 ^2 \Gamma^2 + 4 \alpha_1 ( \rho -\rho_1 ) g }\right ),    \end{align}
        where the plus sign represents right-running waves and the minus sign leads to left-running waves and where 
 \begin{align} \label{Gamma}
            \Gamma:&= \rho \gamma - \rho_1 \gamma_1, \qquad \alpha_1:= \frac{hh_1}{\rho_1h+\rho h_1},
\end{align} are constants.  Therefore, keeping in mind that $\eta_x \sim \varepsilon \delta$ 
\begin{align} \label{Fprim}
    \tilde{F}'(\eta)&= 2\beta \partial^{-1}\eta_t \simeq -2 \beta (\delta \partial) ^{-1}(\varepsilon \delta c_0\eta_x +\mathcal{O}(\varepsilon^2\delta))=-\varepsilon 2\beta c_0\eta+\mathcal{O}(\varepsilon^2) , \\
    \tilde{F}(\eta)& \rightarrow F_0 - \varepsilon^2 \beta c_0 \eta^2 + \mathcal{O}(\varepsilon^3),
\end{align} where $F_0$ is a suitable constant, and, similarly,
\begin{align}
        \tilde{F}_1(\eta)& \rightarrow F_{01} - \varepsilon^2 \beta_1 c_0 \eta^2 + \mathcal{O}(\varepsilon^3).
\end{align} 
Proceeding as in \cite{CoIv2,henry} using the expansions \eqref{dnog}, \eqref{dnog1}, we observe that in the KdV regime the kinematic boundary condition \eqref{kin} gives the same equation, while the Bernoulli equation acquires some new extra terms proportional to $\eta^2$: The term from the RHS of \eqref{Bernoulli} is
\begin{equation}
    \rho \tilde{F}(\eta) - \rho_1 \tilde{F}_1 (\eta) \rightarrow -\varepsilon^2 c_0(\rho \beta - \rho_1 \beta_1)\eta^2 + \frac{(\rho-\rho_1)\kappa^2}{2}.\\
\end{equation} The constant is chosen in such a way as to compensate for the constant arising in the LHS of the equation, see \eqref{LHS} below. The $\eta^2$-contribution from the LHS of \eqref{Bernoulli} is arising from 
\begin{align}
       (\rho U'(\eta) & -\rho_1 U' _1(\eta) )  \partial_x ^{-1}\eta_t  + \frac{\rho U^2(\eta)- \rho_1 U_1^2(\eta)}{2}  
       \nonumber \\
=(\rho \gamma & -\rho_1 \gamma _1 )  \partial_x ^{-1}\eta_t  + 2(\rho \beta - \rho_1 \beta_1) \eta  \partial_x ^{-1} \eta_t     + \frac{(\rho-\rho_1)\kappa^2}{2} +  \frac{(\rho \gamma ^2 -\rho_1 \gamma _1^2  )\eta^2}{2}  \nonumber \\
 & + \kappa (\rho \gamma  -\rho_1 \gamma _1   )\eta + \kappa (\rho \beta - \rho_1 \beta_1) \eta ^2 + \mathcal{O}(\eta ^3)
 \nonumber \\
\rightarrow \varepsilon  \Gamma  \partial_x ^{-1}\eta_t  &  + \varepsilon 2(\rho \beta - \rho_1 \beta_1) \eta [-\varepsilon c_0 \eta +  \mathcal{O}(\varepsilon^2)]    + \frac{(\rho-\rho_1)\kappa^2}{2} + \varepsilon^2 \frac{(\rho \gamma ^2 -\rho_1 \gamma _1^2  )\eta^2}{2}  \nonumber \\
 & + \varepsilon \kappa (\rho \gamma  -\rho_1 \gamma _1   )\eta + \varepsilon^2 \kappa (\rho \beta - \rho_1 \beta_1) \eta ^2 + \mathcal{O}(\varepsilon ^3)
  \nonumber \\
 = \frac{(\rho-\rho_1)\kappa^2}{2}& + \varepsilon [\Gamma  \partial_x ^{-1}\eta_t    + \kappa \Gamma\eta ]  -\varepsilon ^2 (2c_0-\kappa)(\rho \beta - \rho_1 \beta_1) \eta ^2     + \varepsilon^2  \frac{(\rho \gamma ^2 -\rho_1 \gamma _1^2  )\eta^2}{2} + \mathcal{O}(\varepsilon ^3) \label{LHS}
  \end{align}
In terms of the variables $\eta$ and $\mathfrak{u}=\xi_x$ the equations are: 
\begin{align}
\eta_t   &=- [\kappa \eta_x + \alpha_1 \mathfrak{u}_x + \varepsilon \alpha_2 \mathfrak{u}_{xxx} + \varepsilon \alpha_3 (\eta \mathfrak{u})_x+\varepsilon \alpha_4 \eta \eta_x] , \label{BA2Eta} \\
 \mathfrak{u}_t+ &\Gamma \eta_t =  \mathfrak{u}_t+\Gamma f_1 = -[ \kappa \mathfrak{u}_x+ \alpha_5 \eta_x + \varepsilon \alpha_3 \mathfrak{u} \mathfrak{u}_x + \varepsilon \alpha_4 (\eta \mathfrak{u})_x + \varepsilon \alpha_6 \eta \eta_x  ] ,\label{BA2}
\end{align}
where the remaining constants are  
\begin{equation}
    \begin{split}
    \alpha_2&= \frac{h^2h_1^2(\rho h +\rho_1 h_1)}{3(\rho_1 h +\rho h_1 )^2}, \,\,\,\,\, \alpha_3 = \frac{\rho h_1^2- \rho_1 h^2}{(\rho_1 h +\rho h_1)^2}, \quad  \alpha_4 = \frac{\gamma_1 \rho_1 h +\gamma \rho h_1}{\rho_1 h +\rho h_1}, \\
     \alpha _5 & = g(\rho-\rho_1)+\kappa \Gamma ,  \quad   \alpha_6 = \rho\gamma^2-\rho_1\gamma_1 ^2-2c(\rho \beta -\rho_1 \beta_1), \\
     c&=c_0 -\kappa.
    \end{split}
\end{equation}
The equations have a quasi-Hamiltonian structure, described in \cite{CoIv2,henry}. With the Hamiltonian $ H^{(3)}$ in the third order of $\varepsilon$ the equations are
\begin{align}\label{HS}
\eta_t  =&-\left( \frac{\delta H^{(3)}}{\delta \mathfrak{u}} \right)_x , \quad \mathfrak{u}_t+\Gamma \eta_t = -\left( \frac{\delta H^{(3)}}{\delta \eta} \right)_x  ,
\end{align}
\begin{multline} \label{H}
H^{(3)}(\eta,\mathfrak{u})= \varepsilon^2 \frac{1}{2}\alpha_1 \int_{\mathbb{R}}  \mathfrak{u}^2 \, dx 
+\varepsilon^2 \frac{1}{2}\alpha_5 \int_{\mathbb{R}}  \eta^2 \, dx   
+\varepsilon^2 \kappa \int_{\mathbb{R}}\eta  \mathfrak{u} \, dx  \\
-\varepsilon^3 \frac{1}{2}\alpha_2 \int_{\mathbb{R}}  \mathfrak{u}_x^2 \, dx
+ \varepsilon^3 \frac{1}{2}\alpha_3 \int_{\mathbb{R}}\eta  \mathfrak{u}^2 \, dx 
+\varepsilon^3 \frac{1}{2}\alpha_4 \int_{\mathbb{R}}\eta^2  \mathfrak{u} \, dx 
+ \varepsilon^3 \alpha_6 \int_{\mathbb{R}}\frac{\eta^3}{6}  \, dx.
\end{multline}
The one-component reduction to the KdV equation follows the same procedure as in \cite{CoIv2,henry} and gives 
\begin{align}\label{KdV}
 \eta_t &+ c_0 \eta_x  + \varepsilon (  B   \eta_{xxx} +  A   \eta \eta_x)   = \mathcal{O}(\varepsilon^2),  \\
 \mathfrak{u}&=\frac{c}{\alpha_1} \eta+\varepsilon \left( B_1 \eta^2 + B_2 \eta_{xx}\right) + \mathcal{O}(\varepsilon^2),
\end{align}
where 
\begin{align}
A   &  = \frac { 3 \alpha_3  c^2 + 3 \alpha_1  \alpha_4 c + \alpha_6 \alpha_1^2 } { \alpha_1 ( 2 c + \Gamma \alpha_1 ) }  , 
\label{A} \\
    B  & =  \frac {  \alpha_2 c^2 } { \alpha_1 ( 2 c + \Gamma \alpha_1 ) } 
=  \frac 1 3   \frac {  (  \rho h +  \rho_1 h_1 )  c^2 } {  ( 2 c + \Gamma \alpha_1 ) } \alpha_1, \\
\label{Atilde}
B_1 &=  -  \frac {  \alpha_3  c  }{   \alpha_1^2 } -   \frac {   \alpha_4    }{ 2  \alpha_1 } + \frac { 3 \alpha_3  c^2 + 3 \alpha_1  \alpha_4 c + \alpha_6 \alpha_1^2 } { 2 \alpha_1^2 ( 2 c + \Gamma \alpha_1 ) },
\nonumber \\
B_2 &=  \frac { -  \alpha_2  c   (  c + \Gamma \alpha_1  ) }{  \alpha_1^2 ( 2 c +  \Gamma  \alpha_1 )} = - \frac {   \alpha_2  c    }{  \alpha_1^2 } + \frac {   \alpha_2  c^2   }{  \alpha_1^2 ( 2 c +  \Gamma  \alpha_1 ) }.
\end{align}

The limiting case of a single layer with surface waves can be deduced from these results. Since this case is of physical interest, we will present it separately. 

\subsection{ Single layer: $\rho_1=0$}

The single layer case can be obtained formally by setting $\rho_1=0. $  The constant parameters in this case are 
\begin{align}
    \Gamma & \rightarrow \rho \gamma, \quad  \alpha_1 \rightarrow \frac{h}{\rho }, \quad c_0 \rightarrow \kappa +  \frac{1}{2}\left( - \gamma h \pm \sqrt{  \gamma^2 h^2 + 4 gh  }\right ),       \\
    c &= c_0-\kappa= - \frac{ \gamma h}{2} \pm \sqrt{  \frac{\gamma^2 h^2}{4} + gh  } .
\end{align}
Note that $c$ satisfies the equation
\begin{equation}\label{c}
    c^2+\gamma h c - gh =0.
\end{equation}
The remaining limits are
\begin{equation}
    \begin{split}
    \alpha_2& \rightarrow \frac{h^3  }{3 \rho  }, \quad \alpha_3 \rightarrow \frac{1  }{\rho }, \quad  
    \alpha_4 \rightarrow \gamma \quad  \alpha _5 \rightarrow \rho (g +\kappa \gamma) ,  \quad   
    \alpha_6 \rightarrow  \rho(\gamma^2 -2c \beta ) .
    \end{split}
\end{equation}
Using these limits and \eqref{c} we also obtain 
\begin{align}
A   &  \rightarrow  \frac { 3 c^2 + 3 h \gamma c +(\gamma^2 -2c \beta ) h^2  } { h ( 2 c + \gamma h  ) }  =  \frac { 
3 g - 2 c \beta h  +\gamma^2 h   } {  2 c + \gamma h  }  , 
\quad     B   \rightarrow  \frac { h^2 c^2 } { 3  ( 2 c + \gamma h ) } ,
\label{AB} \\
B_1 & \rightarrow  \frac{\rho[3gh-4(c^2 +c\gamma h) -2c\beta h^2]}{2h^2(2c+\gamma h)} = -\frac{\rho (gh +2c\beta h^2)}{2h^2(2c+\gamma h)}
\nonumber \\
B_2 & \rightarrow  -  \frac { \rho  c   h (  c + \gamma h   ) }{ 3  ( 2 c +  \gamma  h  )} 
\end{align}
The limit for the KdV equation therefore is
\begin{equation}\label{kdv1L}
    \eta_t + c_0 \eta_x + \varepsilon \frac{c^2 h^2}{3(2c+\gamma h)} \eta_{xxx}+ \varepsilon 
    \frac{3g- {2}\beta h c + \gamma^2 h}{2c+\gamma h}\eta \eta_x=\mathcal{O}(\varepsilon^2).
\end{equation}
For the fluid potential on the surface we have
\begin{align}
    \mathfrak{u}&=\rho \phi_x (x,t), \\
        \phi_x  & = \frac{ c}{h}\eta  -\varepsilon \left( \frac{gh + {2} c \beta h^2 }{ 2h^2(2c+\gamma h)}\eta^2 + \frac{c h(c+\gamma h)}{3(2c+\gamma h)}  \eta_{xx} \right)+\mathcal{O}(\varepsilon^2) \label{phi}
\end{align}
This case has been studied separately in \cite{Curt2}. 
Here $\phi:=(\varphi)_s=\varphi(x, \eta(x,t),t),$ where subindex $s$ is for the values on the surface. Moreover, by definition, $G(\eta) \phi :=(\varphi_z)_s-(\varphi_x)_s\eta(x).$ Differentiating $\phi$ we have $\phi_x=(\varphi_z)_s \eta_x + (\varphi_x)_s.$ From the last two equations we obtain 
\begin{align}
(\varphi_x)_s& = \frac{\phi_x-\eta_x G(\eta) \phi}{1+\eta_x^2} ,\nonumber \\
(\varphi_z)_s& = \frac{\eta_x\phi_x+ G(\eta) \phi}{1+\eta_x^2} .\nonumber 
\end{align}
In these expressions $\phi$ is given in \eqref{phi} and $G(\eta) $ is given in \eqref{dnog}.

\subsection{Special case $A=0,$ the modified KdV equation }\label{sec:special}

The KdV model \eqref{KdV} is very convenient, since it is integrable and its soliton theory is very well developed, see for example \cite{ZMNP}.  However, this model is problematic in the special case when $A=0,$ or when $A $ is small, for example $A\sim \varepsilon$ or smaller. This situation is possible. For example, in the case without currents, $U(z)=U_1(z) \equiv 0$ the coefficient $A$ is
\begin{equation}
    A=\frac{3c \alpha_3}{2 \alpha_1}= \frac{3c(\rho h_1^2- \rho_1 h^2)}{2\alpha_1(\rho_1 h +\rho h_1)^2}, 
\end{equation} and $A=0$ if the parameters of the system are such that $\rho h_1^2= \rho_1 h^2.$
In this case the dispersive term $\eta_{xxx}$ is balanced by the nonlinear cubic term $\eta^2 \eta_x,$  which requires scaling $\varepsilon \sim \delta. $ We will describe briefly the derivation of the equation in this case of the form
\begin{equation} \label{Gardner}
    \eta_t + c_0 \eta_x + \varepsilon ^2 (B \eta_{xxx}+ A \eta \eta_x + M \eta^2 \eta_x) = \mathcal{O}(\varepsilon^3)
\end{equation} known as Gardner's equation, \cite{Miura,Miura2}. Gardner's equation is integrable, it has stable soliton solutions and has been used extensively in modelling long internal waves, see for example \cite{Gr,HM06} and the references therein.
We note that in this scaling terms of the form $\eta _x \eta_{xx}$ and $\eta \eta_{xxx}$ will appear as $\varepsilon ^3$ terms in the equation and the $\eta_{xxxxx}$ term will be $\varepsilon^4$-contribution. For this reason we do not consider these terms, however, they will contribute in principle to the higher-order KdV-5 equation in the KdV regime, see for example \cite{henry}.

From \eqref{KBC1} with \eqref{dnog}, \eqref{dnog1} we obtain
\begin{align} \label{Eq1}
    \eta_t&=-\left[\kappa \eta_x + \alpha_1 \mathfrak{u}_x + \varepsilon ^2 \alpha_2 \mathfrak{u}_{xxx} + \varepsilon  \alpha_3 (\eta \mathfrak{u})_x+\varepsilon  \alpha_4 \eta \eta_x 
     - \varepsilon ^2 \alpha_7 (\eta^2 \mathfrak{u})_x -
    3 \varepsilon^2 \alpha_8 \eta ^2 \eta_x \right], \\
     \alpha_7&=\frac{\rho \rho_1(h+h_1)^2}{(\rho_1 h + \rho h_1)^3}, \\
    \alpha_8 &:= \frac{\rho \rho_1(\gamma-\gamma_1)( h +h_1)}{2(\rho_1 h + \rho h_1)^2} -\frac{\rho h_1 \beta + \rho_1 h \beta_1}{3(\rho_1 h + \rho h_1)} .
\end{align}

The analogue of \eqref{Fprim} 
\begin{align} \label{Fprim1}
    \tilde{F}'(\eta)&= 2\beta \partial^{-1}\eta_t \simeq -2 \beta   \left [\varepsilon  c_0\eta +\varepsilon^3 \left( B   \eta_{xx} +   \frac{1}{2} A   \eta^2  + \frac{1}{3}  M \eta^3 \right )+ \mathcal{O}(\varepsilon^4)\right]
    \\
    \tilde{F}(\eta)& \rightarrow F_0 - \varepsilon^2 \beta c_0 \eta^2 - \varepsilon^4 \beta \left( B \eta_x^2 + \frac{1}{3} A   \eta^3  + \frac{1}{6} M \eta ^4 \right) + \mathcal{O}(\varepsilon^5),
\end{align} where $F_0$ is the same constant.
The integration of $\eta_{xx} $ can formally be performed as follows: 
\begin{align}
     \int \eta_{xx} d \eta = \int \eta_{xx} \eta_x \, dx = \frac{1}{2} \eta_x^2.
\end{align}
We observe that there is no term of order $\varepsilon^3,$ which is the next order term in the evaluation of quantities, proportional to $\partial^{-1}\eta_t .$  Therefore, taking into account the cubic terms in \eqref{Bernoulli} gives 
\begin{align}
     \mathfrak{u}_t+ \Gamma \eta_t =   -[ \kappa \mathfrak{u}_x &+ \alpha_5 \eta_x + \varepsilon \alpha_3 \mathfrak{u} \mathfrak{u}_x + \varepsilon \alpha_4 (\eta \mathfrak{u})_x + \varepsilon \alpha_6 \eta \eta_x  \nonumber \\
      & -\varepsilon^2 \alpha_7 (\eta \mathfrak{u}^2)_x- \varepsilon ^2 3\alpha_8 (\eta^2  \mathfrak{u})_x 
      -\varepsilon^2 \alpha_9 (\eta^3)_x/2     ] ,       \label{Eq2}      \\
           \alpha_9 & =\frac{\rho\rho_1(\gamma-\gamma_1)^2}{\rho_1 h+\rho h_1}.
\end{align}
The system of equations \eqref{Eq1} and \eqref{Eq2} also has a Hamiltonian structure \eqref{HS} with Hamiltonian
\begin{equation}
    H^{(4)}= H^{(3)} -\varepsilon ^4 \frac{1}{2} \alpha_7 \int_{\mathbb{R}}  \eta^2 \mathfrak{u} ^2 \,dx  - \varepsilon^4 \alpha_8   \int_{\mathbb{R}}  \eta^3  \mathfrak{u} \,  dx  -\varepsilon ^4 \frac{1}{2} \alpha_9 \int_{\mathbb{R}}  \frac{\eta^4}{4} \, dx ,
\end{equation}
where $ H^{(3)}$ is defined in \eqref{H}. We observe that the $\alpha_8$ term is the only term of order $\varepsilon ^4 $ in the Hamiltonian, which depends on $\beta,$ $\beta_1,$ all the other terms are as in the KdV-5 Hamiltonian in \cite{henry}.

The reduction to one-component equation \eqref{Gardner} is with coefficients $A,B$ as before, but $\mathcal{O}(A )\lesssim \varepsilon$ and
\begin{align}
     M & = 3\left [       B_1 \frac { 2 \alpha_3  c  + \alpha_1 \alpha_4   - 4  \alpha_1^2 B_1  } { 3 ( 2 c + \Gamma \alpha_1 ) }  -  \frac { 2 \alpha_7   c^2 } {\alpha_1( 2 c + \Gamma \alpha_1)}  - \frac { 4 \alpha_8 c } { 2 c + \Gamma \alpha_1}  - \frac {  \alpha_9 \alpha_1  } { 2 ( 2 c + \Gamma \alpha_1 ) }  \right]. \label{M}  
 \end{align}
 The relation is 
 \begin{equation}\label{ansatz}
\mathfrak{u}= \frac{c}{\alpha_1} \eta  +\varepsilon B_1 \eta^2 + \varepsilon^2 B_2 \eta_{xx} +\varepsilon^2  B_3 \eta^3 ,
\end{equation}
 \begin{align*}
 B_3  =  -   \frac { \alpha_3 B_1 } {  \alpha_1 } + \frac { \alpha_7  c } { \alpha_1^2 } + \frac {  \alpha_8 } { \alpha_1 }  + \frac { 2  \alpha_1 B_1  \left(  2 \alpha_3 c + \alpha_1 \alpha_4   - 4  \alpha_1^2 B_1 \right)   -   12 \alpha_7   c^2  - 24 \alpha_1 \alpha_8 c   -  3 \alpha_1^2  \alpha_9   } { 6 \alpha_1^2   ( 2 c + \Gamma \alpha_1 ) }.  
\end{align*}

\section{Gardner's equation as a model of equatorial waves} \label{sec:Gardner}

From the considerations in Sub-sections \ref{sec:kdv} and \ref{sec:special} one can conclude that in the general situation with 
$\mathcal{O}(\delta ^2)\lesssim  \mathcal{O}(\varepsilon) \lesssim\mathcal{O}(\delta )$ Gardner's equation 
\begin{equation} \label{Gardner1}
    \eta_t + c_0 \eta_x + \delta ^2 B \eta_{xxx}+ \varepsilon A \eta \eta_x + \varepsilon^2  M \eta^2 \eta_x  = \mathcal{O}(\varepsilon^3, \varepsilon \delta^2, \delta^3)
\end{equation}
can be used, where the constant parameters $c_0,$ $A,$ $B,$ $M$ are previously computed in \eqref{c0}, \eqref{A}, \eqref{Atilde} and \eqref{M}. Discarding the scaling parameters (which only indicate the relative magnitude of the terms), the Gardner's integrable approximate model becomes  
\begin{equation} \label{Gardner2}
    \eta_t + c_0 \eta_x +  B \eta_{xxx}+  A \eta \eta_x +   M \eta^2 \eta_x  = 0.
\end{equation}


The solitons of Gardner's equation depend on its coefficients and other parameters. For example, when $\mathcal{D}= A^2+24 M B K^2 >0,$ there are two one-soliton solutions, given explicitly by
\begin{align}\label{Gardner_sol}
        \eta(x,t)&=\frac{24 B K^2}{A\pm \sqrt{\mathcal{D}}\cosh[2K(x-x_0 - (c_0+4B K^2)t)] }. 
\end{align}
Here, $K$ and $x_0$ are the soliton parameters, that is, arbitrary constants. $K$ is an analog of the wave number and is of order $\delta,$ therefore within the validity of the approximation for small $\delta,$ we usually have $\mathcal{D}>0$ regardless of the sign of $BM .$
In the special case $M=0$ and the plus sign, we have the KdV equation and $\sqrt{\mathcal{D}}=A, $ then we obtain the KdV soliton:
\begin{align}
        \eta_{KdV}(x,t)&=\frac{12 B K^2}{A\cosh^2(K(x-x_0 - (c_0+4BK^2)t)) }.
        \end{align}
The minus sign leads to a KdV soliton $12 BK^2/A\sinh^2(K(x-x_0 - (c_0+4BK^2)t)) $ with singularities when $x-x_0 - (c_0+4BK^2)t=0$. 

A special case is $A=0.$ Then $\mathcal{D}>0$ only if $BM>0$ ( ``focusing'' mKdV) and the soliton becomes
\begin{align}\label{R=0sol}
        \eta(x,t)&=\pm\frac{2\sqrt{6} B K}{\sqrt{BM }\cosh[2K(x-x_0 - (c_0+4BK^2)t)] }.
        \end{align}
    In the situation $A=0,$ $BM<0$ (``defocusing'' mKdV) there is no well-behaved soliton like \eqref{R=0sol}.
However, there is a shock-wave solution 
\begin{equation}\label{sw}
    \eta(x,t)= \pm 2 K \sqrt{-\frac{3B}{M}}\tanh[\sqrt{2}K(x-x_0 - (c_0+4B K^2)t)].
\end{equation}

For the solutions of the Gardner's equation see \cite{Fu}.


We would like to point out that, as demonstrated in the Appendix 2, and also in  \cite{Constantin2021}, a good 
approximation for the EUC involves parameter values $\gamma=\gamma_1=0.$
In this case, the equation parameters for the eastward traveling waves (which are of primary physical interest) are simplified as follows:
\begin{align}
    c_0 =& \kappa+ \sqrt{\alpha_1 (\rho - \rho_1)g}, \quad c=\sqrt{\alpha_1 (\rho - \rho_1)g}, \label{cc}\\
    A =& \frac{ 2(\beta_1\rho_1-\beta \rho)h^2h_1^2 + 3c(\rho h_1^2 - \rho_1 h^2 )}{2h_1 h(h\rho_1 + h_1\rho)} \label{AA}\\
    B =& \frac{h h_1 (h \rho + h_1 \rho_1) c}{6(h\rho_1 + h_1\rho)} , \label{BB}\\
    M =& \frac{\beta_1 h^2\rho_1^2  + \beta h_1^2\rho^2+ \rho\rho_1(\beta_1 h_1^2 + 2 h h_1 (\beta + \beta_1) + h^2\beta)} {(h\rho_1 + h_1 \rho)^2} -\frac{(\beta \rho - \beta_1\rho_1)^2 h^2 h_1^2}{2(h\rho_1 + h_1\rho)^2c}  \nonumber \\
    &-\frac{3(h^4 \rho_1^2+ h_1^4\rho^2) + 6h_1 h \rho \rho_1(4h^2 + 7hh_1 + 4h_1^2) }{8h^2 h_1^2 (h\rho_1 + h_1\rho)^2} c.
\end{align}
Note that in the lack of current ($\beta=\beta_1=0$) we have $M<0,$ and in the critical case when $A=0$ we have the defocusing mKdV equation, which does not possess any soliton solutions.

\section{ Impact of the current on the solitons at the thermocline } \label{sec:7}

\begin{figure}[ht!]
    \centering
\includegraphics[scale=0.250]{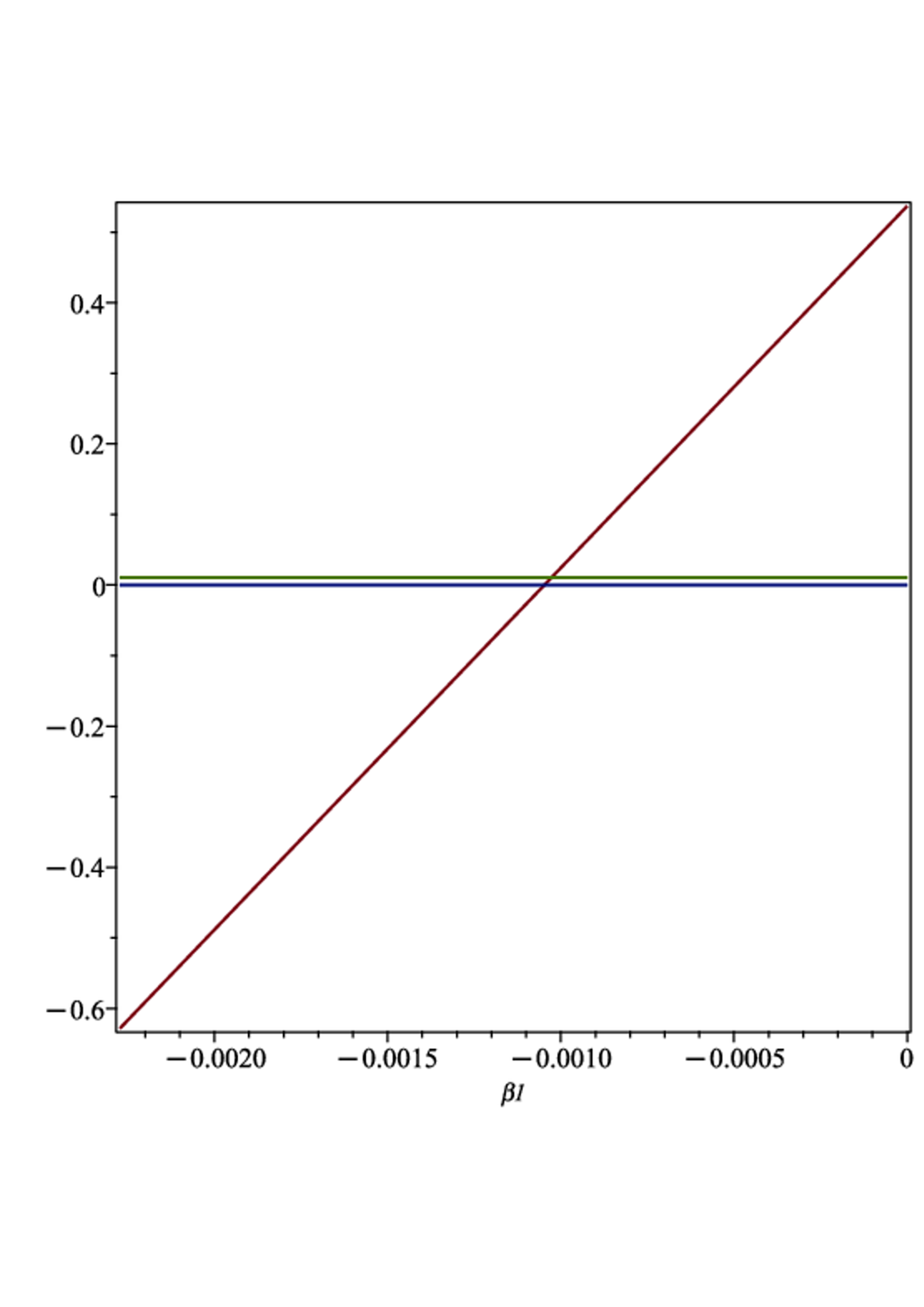}
\vspace{-60pt}
\caption{The coefficient $A$ vs $\beta_1;$ the coefficient $A$ when $\beta=\beta_1=0$ is given by the green horizontal line for comparison; $\rho_1 = 1000.0,$  $\rho = 1027.0,$  $h_1 = 1500.0,$ $h= 800.0.$  }
\label{Fig-beta}
\end{figure}

In this Section we investigate the impact of the current on the solitons forming at the thermocline. For simplicity, we consider the KdV regime, where the effect is most obvious. We use the formulas from the previous Section for the right-running waves \eqref{cc}--\eqref{BB}, where  $c>0,$  $B>0$ and $A$ is given by \eqref{AA}. We note that only the parameter $A$ depends on the current parameters, so let us investigate this dependence. We use SI units for all physical quantities involved. In the Appendix 2 we have estimated $\beta_1 \simeq - 2.27 \times 10^{-3},$ so let us assume a fixed value $\beta = -1.0 \times 10^{-3} $ of the same magnitude. Then we vary $\beta_1$ in the interval $[- 2.27 \times 10^{-3},\,\, 0.0].$ For the other parameters we also take the very typical values: $\rho_1 = 1000.0,$
 $\rho = 1027.0,$  $h_1 = 1500.0,$ $h= 800.0,$ ($c = 11.65$). As a result, we obtain the linear dependence $A(\beta_1)$ presented in Fig. \ref{Fig-beta}. For comparison, the value $A$ when there is no current $\beta=\beta_1=0$ is given in green. 
 As a final result we observe that the nonlinearity coefficient $A$ is highly sensitive with respect to the current parameters (in this case $\beta_1$). Moreover, since the soliton polarity depends on the sign of $A$ we see that the current can change the soliton polarity of a soliton when all other physical parameters are the same. This is an example for a change of the soliton polarity due to currents. (Another cause of the polarity conversion could be tides, which can alter the geometry of the system - the $h_1/h$ ratio, as reported in \cite{LWG}.)
 
   This example also shows that the analysis of any data about the solitary waves on the thermocline in the Pacific is not possible (or, is not correct) without taking into account the data about the parameters of the current. 

\section{Internal waves over variable bottom} \label{sec:8}

The effect of the variable bottom can be studied following the methods in \cite{IMT}. 
Let us assume that the variable bottom is located at $z=-h + \beta(x), $ that is, the function $\beta(x)$ describes the bottom variations. The bottom variations are also considered small, but $|\beta_{\mathrm{max}}|/h $ of order $\tilde{\epsilon} \le \delta^{2/3}.$
The rationale of this choice is explained in \cite{IMT} in the case when $\varepsilon$ and $\delta^2$ are of the same order. The other assumption is related to the slow bottom variations. 
Mathematically, this assumption is given as $\beta= \beta(\delta^2 x).$ Then the commutator of $\beta$ and the differentiation operator $D:=-i\partial_x$ is proportional to $\delta^2 \beta'(\delta^2 x)$ which itself is of order $\delta^2$. Introducing the local depth as $b(X)= h-\beta(\delta^2 x)$ where $X=\delta^2 x$ indicates that the bottom depth varies slowly with $x$, the DNO \eqref{dnog} and \eqref{dnog1} acquire the form
\begin{equation}
\begin{split}
G(b, \eta)&= \delta^2 D( b(X)+\varepsilon \eta)D -\delta^4 D^2 \left[ \frac{1}{3}b^3(X)+\varepsilon h^2 \eta \right]D^2 + \delta^6 \frac{2}{15}h^5 D^6+\mathcal{O}(\delta^8, \varepsilon \delta^6, \varepsilon^2 \delta^4) , \label{DN_1}
\end{split}
\end{equation}
\begin{equation}\label{G1}
G_1(\eta)=\delta^2 D\left( h_1 -\varepsilon  \eta \right)D -\delta^4 D^2  \left[ \frac{1}{3}h_1^3-\varepsilon h_1^2 \eta \right]D^2   +\mathcal{O}(\delta^8, \varepsilon \delta^6, \varepsilon^2 \delta^4).
\end{equation}

As a result of the slowly variable bottom all parameters become slowly varying functions of $X,$ and can be obtained by replacing the average depth $h$ by the local depth $b(X),$ for example
\begin{align} \label{c0alpha1}
\alpha_1(X)&= \frac{b(X)h_1}{\rho_1 b(X)+\rho h_1}, \quad c(X)= c_0(X) - \kappa, \\
        c_0 (X)&= \kappa +  \frac{1}{2}\left( - \alpha_1(X) \Gamma \pm \sqrt{ (\alpha_1(X)) ^2 \Gamma^2 + 4 \alpha_1(X) ( \rho -\rho_1 ) g }\right ), \, \, \, \text{etc.}    \end{align}

The computations follow the same pattern, using the generalised version of the relation \eqref{ansatz} 
 \begin{equation}\label{ansatz1}
\mathfrak{u}= \frac{c}{\alpha_1} \eta + \delta^2 f(X) \partial_x^{-1}\eta   +\varepsilon B_1(X) \eta^2 + \varepsilon^2 B_2 (X)\eta_{xx} +\varepsilon^2  B_3 (X) \eta^3 ,
\end{equation}  where, the unknown slowly varying function $f$ can be obtained in the same way as the other coefficients $B_k(X),$  that is,
\begin{equation}
    f(X)=-  \frac{ [c(X)+\alpha_1(X) \Gamma ][c(X)+\kappa] c'(X)}{\alpha_1(X) c(X) [2c(X)+\alpha_1(X) \Gamma ]}.
\end{equation}
   
Gardner's equation \eqref{Gardner1} generalises to an equation with slowly varying coefficients
\begin{align} 
    \eta_t +& c_0(X) \eta_x + \delta^2   \frac{ [c^2 -\kappa \big(c(X)+\alpha_1(X) \Gamma \big)] c'(X)}{ c(X) [2c(X)+\alpha_1(X) \Gamma ]}     \eta  \nonumber \\
    &+ \delta ^2 B(X) \eta_{xxx}+ \varepsilon A(X) \eta \eta_x + \varepsilon^2  M(X) \eta^2 \eta_x  = \mathcal{O}(\varepsilon^3, \varepsilon \delta^2, \delta^3)        \label{Gardner2}
\end{align}

The soliton propagation over variable bottom leads to effects such as soliton fission,
which are analysed and studied numerically in \cite{IMT} in slightly different coordinates and under the assumption $\beta=\beta_1=0.$

In \cite{GPT} the solitons of the Gardner's equation are presented and analysed in detail. The variable coefficients Gardner's equation is reported (in the irrotational case $\gamma=\gamma_1=0,$ $\beta=\beta_1=0$) as a model of the  internal solitary waves over topography in the coastal oceans in different geographic areas such as three representative continental shelves, the North West Shelf (NWS) of Australia, the Malin Shelf off the North West coast of Scotland, and the Laptev Sea in the Arctic. The soliton fission is discussed as well as the variability of the equation coefficients with $X.$ 

In some cases the KdV with slowly varying coefficients is equivalent to forced KdV. Thus, with this mechanism, trapped interfacial waves may arise through resonance with the bottom topography, as demonstrated in \cite{F1,F2}.

\section{Conclusions}


The presented approach is in a sense an alternative to the approach, based on the Burns condition \cite{Burns,Johnson}, which has been applied to internal waves in \cite{GQ}.  The Burns condition for a single layer determines the leading order of the wave speed in the small amplitude approximation via the integral over the entire fluid depth 
\begin{equation} \label{BC}
    \int_{-h}^{0} \frac{dz}{\left(U(z) - c\right)^2}.
\end{equation}
Thus, for some functions $U(z)$ the integral can encounter singularities whenever $U(z) = c$ for some $z$. 
The detailed analysis shows that in the ``small amplitude'' assumption for the wave motion,  $c$ depends only on the linear part of $U(z),$ that is, in the single layer case $c $ depends on $\kappa$ and $\gamma$ only, while the quadratic part of $U(z)$ affects only the nonlinear part of the model equations - this is evident from our analysis here and in \cite{Curt2}. Moreover, one can show that in the case of linear current, Burns condition produces the same wavespeed $c,$ see for example \cite{Iv07}. 

Therefore, since the Burns condition relies on the small amplitude approximation, it is more appropriate to use the linearised part of the current on the surface/interface, that is, to use $U_{L}(z)=\kappa+\gamma z$ instead of $U(z)$ in \eqref{BC}. Then, indeed, there is no critical layer phenomenon when the function $U(z)\equiv U_{L}(z)$ is linear. Moreover, as it was shown here and in \cite{Curt2}, $c$ is determined only by the linear component of the current $U_L.$ 

Both the KdV and Gardner's equations have the huge advantage that they are integrable nonlinear equations and the soliton theory for such equations is very well developed. In the framework of the soliton theory one can investigate the propagation of solitary waves but also other types of solutions.

It is also possible to obtain higher order KdV models, including some integrable ones, following the techniques from \cite{henry} and including further terms with nonlinearities and dispersion. However, for practical applications, Gardner's equation is the golden standard as an integrable nonlinear prototype model for internal waves.  

The coefficients of the model \eqref{Gardner1} and its soliton solutions, which are the most stable structures in time and amenable to detection and observation, depend substantially on the current parameters. This means that any field data for the motion of the thermocline can be properly understood only if they are supplemented by the data about the EUC (e.g. the parameters $\kappa,$ $\gamma,$ $\gamma_1,$ $\beta,$ $\beta_1$) and only if these parameters are taken into account. It is evident that the role of the EUC goes far beyond a simple "translation" of the thermocline and affects both the linear and nonlinear aspects of the wave propagation. 

The main interest of the study has been on the motion of the thermocline. However, the velocity field can be extended in the fluid domains by analyticity due to equalities of the type \eqref{CR}. Then one interesting aspect of considering an underlying shear current is the possible appearance of stagnation points within the fluid beneath the solitons. Such scenarios have been investigated recently in \cite{F3,F4}. For instance, in \cite{F3}, the authors demonstrated the existence of stagnation points for various underlying shear currents; for example for large $\gamma$ and $\gamma_1$ for linear shared currents. It needs to be checked however if these assumptions are compatible with the assumptions about the validity of the Gardner's equation.

\subsection*{Data availability statement }

The reported results are of purely theoretical nature, and no data have been used or generated to support these results.

\subsection*{Acknowledgments} This publication has emanated from research conducted with the financial support of Taighde \' Eireann – Research Ireland under Grant number 21/FFP-A/9150. The authors are thankful to two anonymous referees for their constructive comments and suggestions. 

\subsection*{Author Contributions}

C.T.C. and R.I.I. contributed equally to this work. C.T.C.: Methodology, Formal analysis, Visualization,
Writing – original draft, Writing – review \& editing; R.I.I.: Conceptualization, Methodology, Formal analysis, Funding acquisition, Writing – original draft, Writing – review \& editing.

\section{Appendix 1}

In this appendix we analyse the effect of the Coriolis forces on the equations of motion. The Coriolis forces are taken into account explicitly in the previous work \cite{CoIv2}, which studies the internal waves with currents in the KdV regime. The first observation is that the Coriolis forces (${\bf F} =2\Omega_E \nabla \psi$ in the domain $\Omega$, ${\bf F} _1=2\Omega_E \nabla \psi _1$ in the domain $\Omega_1$) affect only the linear propagation regime of the system and modify the constant $\Gamma$ in \eqref{Gamma} as follows
 \begin{align} \label{Gamma1}
            \Gamma:&= \rho \gamma - \rho_1 \gamma_1
            +2\Omega_E (\rho- \rho_1),
\end{align} where $\Omega_E= 7.292 \times 10^{-5}$ radians per second is the angular speed of the Earth rotation. The equation for $c$ also depends on $\Gamma,$
\begin{equation}\label{eq4c}
    c^2 + \alpha_1 \Gamma c - \alpha_1 (\rho- \rho_1) g =0
\end{equation} and therefore $c$ depends on the Coriolis force as well. Let us make an estimation of the magnitude of this dependence. Let us take $0.01$ s$^{-1}$ as a typical value for $\gamma_1$ - this corresponds to  surface speed due to wind of several m/s, decreasing to zero at the thermocline over the top several hundred meters and $\gamma=0$ for the lower layer.
Thus $\rho \gamma -\rho_1\gamma_1 = 0-0.01\times 1000=-10^1.$ On the other hand, the maximum density difference is no more than $30$ kg/ m$^3$, so the Coriolis contribution is  
$2\Omega_E (\rho- \rho_1)\simeq 2\times  7.292 \times 10^{-5}\times 30 \simeq 4\times 10^{-3},$ which is several orders smaller. Let us now examine the effect on $c.$ Let us consider a small variation $\delta c$ due to the Coriolis force, assuming $\delta \Gamma= 2\Omega_E (\rho- \rho_1).$  From \eqref{eq4c} we have
\begin{equation}
    \frac{\delta c}{c} = -\frac{\alpha_1}{2c+\alpha_1 \Gamma}\delta \Gamma 
\end{equation}
Assuming further an average $\Gamma=0,$ $h=h_1=500$ m,  density difference $30$ kg/ m$^3$,  we have $c\simeq 3.7$ m/s which is indeed in the range of the expected typical values (typically detected in the ocean).  Then 
\begin{equation}
    \frac{\delta c}{c} \lesssim 2.7 \times 10^{-5}.
\end{equation}
Therefore, the influence of the Coriolis forces on the parameters $\Gamma$ and $c$ is small. We should not forget however, that the Coriolis forces at the equator play another important role by supporting the flow to remain essentially 2-dimensional along the equator. 
And if necessary, one can use the more precise value for $\Gamma,$ given by \eqref{Gamma1}.

\section{Appendix 2} 

In this appendix, we describe a simplified framework for the derivation of the wind-induced current velocity field in the equatorial region of the Pacific ocean. We ignore some circumstances, taking into account only the factors that play the leading role in the formation of the phenomena being studied. For example, we assume the wind stress on the surface to be constant, and we consider the linearized equations for a forced steady-state flow with a vanishing vertical fluid velocity.  
To some extent we follow the derivation  in \cite{Constantin2021,CJ}.

We are looking for a steady velocity field solution of the of the linearised Navier-Stokes equations in the $f$-plane approximation on the Equator, representing current in the form $(U_1(z),0,0) $ in the upper layer ($0\le z \le h_1$) and $(U(z),0,0) $ in the lower layer ($-h \le z \le 0$). The wind stress on the surface determines the constant value $U_1(h_1)=U^*$ on the surface, which is usually negative - in the westward direction.  We assume that the pressure $P(x,z)$ depends on both coordinates. The mass conservation is trivially satisfied. The quantities in the upper layer, as usual, bear subindex 1.  The Navier-Stokes equations, with the Coriolis force included, are
\begin{align}
    0&= -\frac{1}{\rho_1}P_{1,x} + \nu_1 U_1 ''(z) , \\
    -2 \Omega_E U_1 (z) &= -\frac{1}{\rho_1}P_{1,z} - g
\end{align}
for the upper layer, where $\Omega_E$ is the angular speed of the Earth rotation and $\nu_1$ is the vertical eddy viscosity of the upper layer, which, for simplicity is assumed constant\footnote{The value of the vertical eddy viscosity is several orders of magnitude larger than the molecular kinematic viscosity of water, which is roughly $1\times 10^{-6}$ m$^2$/s at room temperature.}.
Moreover, since the current is of magnitude of several m/s we have $\Omega_E U_1(z) \ll g $ and the Coriolis term can be neglected, thus 
\begin{align}
    0&= -\frac{1}{\rho_1}P_{1,x} + \nu_1 U_1 ''(z) , \label{NS1}\\
   0 &= -\frac{1}{\rho_1}P_{1,z} - g, \label{NS2}
\end{align} and similarly, for the lower layer
\begin{align}
    0&= -\frac{1}{\rho}P_{x} + \nu U ''(z) ,  \label{NS3} \\
    0 &= -\frac{1}{\rho}P_{z} - g. \label{NS4}
\end{align}
The relevant boundary conditions are as follows:\\
(i) on the free surface $z=h_1$,  the wind stress condition 
$$\tau_1= \rho_1 \nu_1 U_{1}'(h_1) $$
(ii) The shear is continuous, on the thermocline $$U_{1}(0) = U(0) , $$
and that the horizontal velocity component and the pressure are continuous throughout the thermocline $z = 0,$ \\
(iii) $U(-d_0)=0$ for some $d_0< h$ expressing the condition that the currents are confined in the layer above the level 
$z=-d_0$  so that the water is still below this level.

From \eqref{NS2} we have $P_1(x,z)=P_{10}(x) -\rho_1 g z,$ from \eqref{NS1} and the boundary conditions we have 
\begin{align}
        \nu_1 U_1(z)&= \nu_1 (U_{10}+U_{11}z) +\frac{P_{10}'(x)}{2 \rho_1} z^2, \quad  U_{1k} = \text{const}. \\
        \tau_1 & = \rho_1  \nu_1 U_1'(h_1)= {P_{10}'(x)} h_1 + \rho_1 \nu_1 U_{11},
\end{align}
Moreover, the solution makes sense only if the horizontal pressure gradient $$ a:= P_{10}'(x)= \text{const. }$$
Then 
\begin{align}
 U_{11} & = \frac{\tau_1  -  a h_1 }{\rho_1 \nu _1}, \\
        \nu_1 U_1(h_1)&= \nu_1 \left(U_{10} +\frac{\tau_1  -  a h_1 }{\rho_1 \nu_1 } h_1  \right) +\frac{a}{2 \rho_1} h_1^2, 
        \\
         U_{10} & =  U_1(h_1) - \frac{\tau_1   h_1 }{\rho_1 \nu_1}     + \frac{a h_1^2}{2\nu_1  \rho_1}, \\
        U_1(z)&=   U_1(h_1) + \frac{\tau_1  -  a h_1 }{\rho_1 \nu_1}(z- h_1)   + \frac{a}{2\nu_1  \rho_1} (z^2 - h_1^2)
        \nonumber \\
        &\equiv U^* +  \frac{\tau_1  -  a h_1 }{\rho_1 \nu_1}(z- h_1)+\frac{a}{2\nu_1  \rho_1} (z^2 - h_1^2). \label{coeff1}
\end{align}
From \eqref{NS4} we have $P(x,z)=P_0(x) -\rho g z,$ the equality of the two pressures at the interface $P_1(x,0)=P(x,0),$
thus $ P_{10}(x)= P_0(x).$ From \eqref{NS3} we have 
\begin{equation}
     U(z)= \kappa + \gamma z +\frac{a}{2 \nu \rho} z^2, \qquad \kappa, \gamma =\text{const.}
    \end{equation} 
The continuity on $z=0$ gives
\begin{equation}
 \kappa  =   U^* -  \frac{\tau_1 h_1 }{\rho_1 \nu_1}  +\frac{ah_1^2}{2\nu_1  \rho_1} .
\end{equation}
and $U(-d_0)=0$    gives 
\begin{align}
    0&= \kappa -\gamma d_0 +\frac{a}{2 \nu \rho} d_0^2, \\
  \gamma &= \frac{1}{d_0}  \kappa +\frac{a}{2 \nu \rho} d_0=\frac{U^*}{d_0} -  \frac{\tau_1  h_1 }{\rho_1 \nu_1 d_0}  +\frac{ah_1^2}{2\nu_1  \rho_1 d_0}    +\frac{a d_0}{2 \nu \rho} .
  \end{align}
Setting $$\beta:=\frac{a}{2 \nu \rho} $$ for $U(z) $ we obtain the quadratic function $U(z)= \kappa+ \gamma z + \beta z^2.$
We note that $U_1(z) $ is also a quadratic function $U_1(z)= \kappa+ \gamma_1 z + \beta_1 z^2$ where, from \eqref{coeff1} 
\begin{align}
    \beta_1 &=\frac{a}{2\nu_1  \rho_1},  \label{beta1}\\
    \gamma_1 &=\frac{\tau_1  -  a h_1 }{\rho_1 \nu_1}.
\end{align}

In some models a rigid bottom condition is further imposed on the top layer, or even on both layers \cite{Constantin2021}. In the case of a rigid bottom for the top layer, we have the additional approximation
\begin{equation}
    U_{1,z}(0)=U_1'(0)=0, \,\, \text{or}\,\, \gamma_1=0,
\end{equation} 
giving also the relation between wind stress and horizontal pressure gradient 
\begin{equation} \label{tau1}
    \tau_1  =  a h_1.
\end{equation} 
Since the winds on the surface are in a westward direction, $\tau_1<0,$ $a<0.$

Then geometrically, the vertex of the parabola $U_1(z)=\kappa+\beta_1 z^2$ is located at $z=0.$
If the same condition is imposed on the lower layer as well, $U'(0)=0,$
\begin{align}
      \gamma &= \frac{U^*}{d_0} -  \frac{ah_1^2}{2\nu_1  \rho_1 d_0}    +\frac{a d_0}{2 \nu \rho}=0  \\
         d_0^2 &=    \frac{ \nu \rho}{\nu_1  \rho_1 } {h_1^2} - \frac{2 \nu \rho}{a}  U^*  \label{h1}
  \end{align}
In the equatorial setting, both $a$ and $  U^* $ are negative, so the model is adequate only if 
\begin{equation} \label{h1a}
    h_1^2 >  \frac{2 \nu_1 \rho_1}{|a|}  |U^*| \, \, \, \text{or } \, \,  h_1 >  \frac{2 \nu_1 \rho_1}{|\tau_1|}  |U^*|=:h^* .
    \end{equation}
The typical data figures : The vertical eddy viscosity in the surface mixed layer is in the range of  $\nu_1 \sim 2.2\times 10^{-4}$ m$^2$/s to  $\nu_1 \sim 13.4 \times 10^{-4}$ m$^2$/s,   $\rho_1 \sim 10^3$ kg/m$^3$, $|\tau_1|\sim 0.1 $ Pa =$0.1$ N/m$^2$ - see for example \cite{Sto,Sen}; $|U^*| \lesssim 1.5$ m/s, giving
$$ h^*:=\frac{2 \nu_1 \rho_1}{|\tau_1|}  |U^*| \lesssim  40 \, \text{m}.$$ This is usually less than $h_1,$ which is of magnitude tens to hundreds of meters. Thus, $d_0$ and $h_1$ are quantities of the same magnitude. Since the lower layer is usually deeper than the upper layer, in most realistic scenarios $h>d_0,$ however, the other possibility is also possible, in which case the distance $d_0$ is below the bottom and the current at the bottom is nonzero.

We also notice that in the case where $\gamma = \gamma_1 =0$ the vertex of both parabolas $U(z)=\kappa + \beta z^2$ and $U_1(z)=\kappa + \beta_1 z^2$ co-inside and are on the thermocline $z=0,$ see Fig. \ref{FigU}. Then
\begin{equation}
U_{max}= \kappa  =   U^* -  \frac{\tau_1 h_1 }{2\rho_1 \nu_1} =U^* \left( 1- \frac{ h_1 }{h^*}\right) . \nonumber
\end{equation}
Under condition \eqref{h1a} bearing in mind that $ U^* <0$ the current on the thermocline is eastward ($\kappa>0$) with typicl values $\kappa \lesssim 1$ m/s.  

 From  \eqref{beta1}  and \eqref{tau1} we can estimate the value of $\beta_1,$ assuming
$\nu_1 \simeq  2.2\times 10^{-4}$ m$^2$/s,   $\rho_1 \simeq 10^3$ kg/m$^3$, $\tau_1\simeq - 0.1 $ Pa, $h_1 = 100$ m  :
\begin{equation}
    \beta_1=\frac{\tau_1}{2\nu_1  \rho_1 h_1 } \simeq - 2.27 \times 10^{-3}  \text{m}^{-1} \, \, \text{s}^{-1}.  \label{beta1-est}
\end{equation}

In this appendix we have used many simplifying assumptions about the EUC. For a deeper analysis we refer to \cite{CJ} and \cite{Shul}. The explicit formula for the current in \cite{Shul} depends on more parameters but is also piecewise-quadratic in $z.$

\end{document}